\documentclass[sigconf]{acmart}

\usepackage{graphicx}
\usepackage{courier}
\usepackage{url}
\usepackage{subcaption}
\usepackage{booktabs}
\usepackage{amssymb}
\usepackage{pifont}
\newcommand{\cmark}{\ding{51}}%
\newcommand{\xmark}{\ding{55}}%
\usepackage{multirow}

\usepackage{color}

\AtBeginDocument{%
  \providecommand\BibTeX{{%
    \normalfont B\kern-0.5em{\scshape i\kern-0.25em b}\kern-0.8em\TeX}}}

\setcopyright{acmcopyright}
\copyrightyear{2020}
\acmYear{2020}
\acmDOI{10.1145/3394171.3413671}

\acmConference[MM '20] {28th ACM International Conference on Multimedia}{October 12--16, 2020}{Seattle, WA, USA.}
\acmBooktitle{28th ACM International Conference on Multimedia (MM '20), October 12--16, 2020, Seattle, WA, USA.}
\acmPrice{15.00}
\acmDOI{10.1145/3394171.3413671}
\acmISBN{978-1-4503-7988-5/20/10}

\begin{document}

\fancyhead{}

\title{Pop Music Transformer: Beat-based Modeling and Generation of Expressive Pop Piano Compositions}


\author{Yu-Siang Huang}
\affiliation{%
  \institution{Taiwan AI Labs \& Academia Sinica }
  \city{Taipei}
  \country{Taiwan}
}
\email{yshuang@ailabs.tw}

\author{Yi-Hsuan Yang}
\affiliation{%
  \institution{Taiwan AI Labs \& Academia Sinica}
  \city{Taipei}
  \country{Taiwan}
}
\email{yhyang@ailabs.tw}

\renewcommand{\shortauthors}{Yu-Siang Huang and Yi-Hsuan Yang}

\begin{abstract}
A great number of deep learning based models have been recently proposed for automatic music composition.  Among these models, the Transformer stands out as a prominent approach for generating expressive classical piano performance with a coherent structure of up to one minute. The model is powerful in that it learns abstractions of data on its own, without much human-imposed domain knowledge or constraints. In contrast with this general approach, this paper shows that Transformers can do even better for music modeling, when we improve the way a musical score is converted into the data fed to a Transformer model. In particular, we seek to impose a metrical structure in the input data, so that Transformers can be more easily aware of the beat-bar-phrase hierarchical structure in music. The new data representation maintains the flexibility of local tempo changes, and provides hurdles to control the rhythmic and harmonic structure of music. With this approach, we build a Pop Music Transformer that composes Pop piano music with better rhythmic structure than existing Transformer models. 
\end{abstract}



\maketitle

\section{Introduction}
\label{sc:introduction}
Music is sound that's organized on purpose on many time and frequency levels to express different ideas and emotions. For example, the organization of musical notes of different fundamental frequencies (from low to high) influences the melody, harmony and texture of music. The placement of strong and weak beats over time, on the other hand, gives rise to the perception of rhythm \cite{martineau2008elements}. Repetition and long-term structure are also important factors that make a musical piece coherent and understandable.

\begin{table}[!t]
\centering
\begin{tabular}{@{}l|ll@{}}
\toprule
 & MIDI-like \cite{oore2018time} & REMI (this paper) \\ \midrule
Note onset & \begin{tabular}[c]{@{}l@{}}\textsc{Note-On}\\ (0--127)\end{tabular} & \begin{tabular}[c]{@{}l@{}}\textsc{Note-On}\\ (0--127)\end{tabular} \\ \midrule
Note offset & \begin{tabular}[c]{@{}l@{}}\textsc{Note-Off}\\ (0--127)\end{tabular} & \begin{tabular}[c]{@{}l@{}}\textsc{Note Duration}\\ (32th note multiples; 1--64)\end{tabular} \\ \midrule
Time grid & \begin{tabular}[c]{@{}l@{}}\textsc{Time-Shift}\\ (10--1000ms)\end{tabular} & \begin{tabular}[c]{@{}l@{}}\textsc{Position} (16 bins; 1--16)\\ \&~\textsc{Bar} (1)  \end{tabular} \\ \midrule
Tempo changes & \xmark & \begin{tabular}[c]{@{}l@{}}\textsc{Tempo}\\ (30--209 BPM)\end{tabular} \\ \midrule
Chord & \xmark & \begin{tabular}[c]{@{}l@{}}\textsc{Chord}\\ (60 types)\end{tabular} \\ \bottomrule
\end{tabular}
\caption{The commonly-used MIDI-like event representation \cite{oore2018time,huang2018music} versus the proposed beat-based one, REMI. In the brackets, we show the ranges of the type of event.}
\label{tab:event-comparison}
\end{table}

Building machines that can compose music like human beings is one of the most exciting tasks in multimedia and 
artificial intelligence \cite{bottoni06mm,yu12mm,lyu15mm,pachet16tist,gatti17mm,liu17survey,chen18mm,xiaoice18kdd,musegan}.
Among the approaches that have been studied, neural sequence models, which consider music as a \emph{language}, stand out in recent work \cite{oore2018time,huang2018music,payne2019musenet,choi2019encoding} as a prominent approach with great potential.\footnote{However, whether music and language are related is actually debatable from a musicology point of view. The computational approach of modeling music in the same way as modeling natural language therefore may have limitations.}
In doing so, a digital representation of a musical score is converted into a time-ordered sequence of discrete tokens such as the \textsc{Note-On} events.
Sequence models such as the Transformer \cite{vaswani2017attention} can then be applied to model the probability distribution of the event sequences, and to sample from the distribution to generate new music compositions.
This approach has been shown to work well for composing minute-long pieces of classical piano music with expressive variations in note density and velocity \cite{huang2018music}, in the format of a MIDI file.\footnote{Therefore, the model generates the pitch, velocity (dynamics), onset and offset time of each note, including those involved in the melody line, the underlying chord progression, and the bass line, etc in an expressive piano performance.}

We note that there are two critical elements in the aforementioned approach---the way music is converted into discrete tokens for language modeling, and the machine learning algorithm used to build the model. Recent years have witnessed great progress regarding the second element, for example, by improving the self-attention in Transformers with ``sparse attention'' \cite{child2019generating}, or introducing a recurrence mechanism as in Transformer-XL for learning longer-range dependency \cite{dai2019transformer,donahue2019lakhnes}.
However, no much has been done for the first one. Most work simply follows the \emph{MIDI-like} event representation proposed by \cite{oore2018time} to set up the ``vocabulary'' for music.  

As shown in Table \ref{tab:event-comparison}, the MIDI-like representation, for the case of modeling classical piano music \cite{huang2018music,choi2019encoding}, uses \textsc{Note-On} events to indicate the action of hitting a specific key of the piano keyboard, and \textsc{Note-Off} for the release of the key.
This representation has its roots in MIDI \cite{heckroth1998tutorial}, a communication protocol that also uses \textsc{Note-On} and \textsc{Note-Off} messages to transmit data of real-time musical performance. 
Unlike \emph{words} in human language, note messages in MIDI are associated with \emph{time}. To convert a score into a sequence, \cite{oore2018time} uses additionally the \textsc{Time-Shift} ($\Delta T$)
events to indicate the relative time gap between events (rather than the absolute time of the events), thereby representing the advance in  time. 
This MIDI-like representation is general and represents keyboard-style music (such as piano music) fairly faithfully, therefore a good choice as the format for the input data fed to Transformers.
A general idea here is that a deep network can learn abstractions of the MIDI data that contribute to the generative musical modeling on its own. Hence, high-level (semantic) information of music \cite{bogdanov09ism}, such as downbeat, tempo and chord, are not included in this MIDI-like representation.

%
\begin{figure}[!t]
\centering
\begin{subfigure}[t]{\linewidth}
    \centering
    \includegraphics[width=\linewidth]{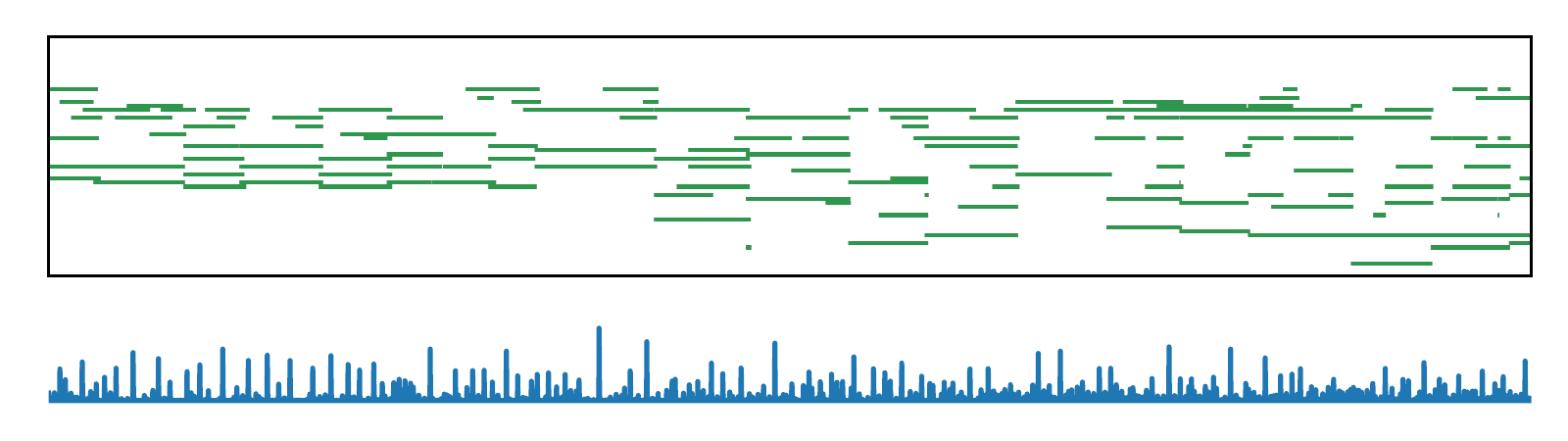}
    \caption{Transformer-XL $\times$ MIDI-like (`Baseline 1')}
    \label{fig:salience:MIDI}
\end{subfigure}
\par\medskip
\begin{subfigure}[t]{\linewidth}
    \centering
    \includegraphics[width=\linewidth]{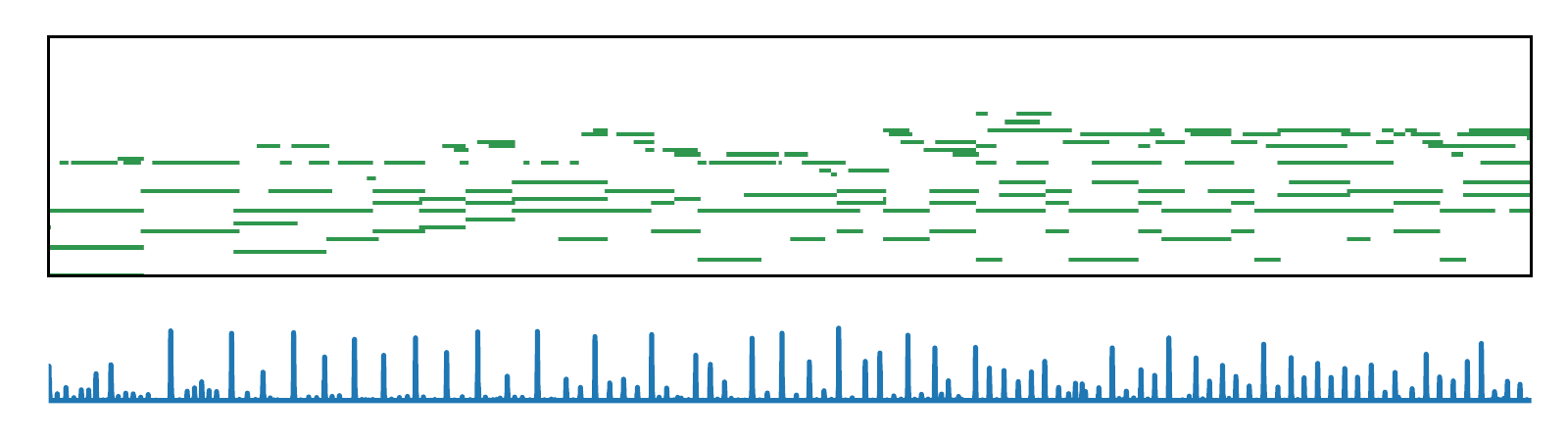}
    \caption{Transformer-XL $\times$ REMI (with \textsc{Tempo} and \textsc{Chord})}
    \label{fig:salience:REMI}
\end{subfigure}
\caption{Examples of piano rolls and  `downbeat probability curves' (cf. Section \ref{ssc:objective-evaluation}) for music generated by an adaptation of the state-of-the-art model Music Transformer \cite{huang2018music}, and the proposed model. We can see clearer presence of regularly-spaced downbeats in the probability curve of (b). }
\label{fig:salience}
\end{figure}

However, we note that when humans compose music, we tend to organize regularly recurring patterns and accents over a \emph{metrical structure} defined in terms of sub-beats, beats, and bars \cite{cooper1963rhythmic}. Such a structure is made clear on a score sheet or a MIDI file with notation of the time signature and bar lines, but is \emph{implicit} in the MIDI-like event representation. A sequence model has to recover the metrical structure on its own from the provided sequence of tokens. When it comes to modeling music genres that feature the use of steady beats, we observe that the  Transformer \cite{huang2018music} has a hard time learning the regularity of beats over time, as shown in Figure \ref{fig:salience:MIDI}. 

To remedy this, and to study whether Transformer-based composition models can benefit from the addition of some human knowledge of music, we propose \emph{REMI}, which stands for \underline{re}vamped \underline{MI}DI-derived events, to represent MIDI data following the way humans read them.
Specifically, we introduce the \textsc{Bar} event to indicate the beginning of a bar, and the \textsc{Position} events to point to certain locations within a bar.
For example, \textsc{Position (9/16)} indicates that we are pointing to the middle of a bar, which is quantized to 16 regions in this implementation (see Section \ref{sc:event-based-representation} for details).
The combination of \textsc{Position} and \textsc{Bar} therefore provides an \emph{explicit} metrical grid to model music, in a beat-based manner. This new data representation informs models the presence of a beat-bar hierarchical structure in music and empirically leads to better rhythmic regularity in the case of modeling Pop music, as shown in Figure \ref{fig:salience:REMI}.

We note that, while the incorporation of the bars is not new in the literature of automatic music composition \cite{Sturm2016a,Hadjeres2016a,briot19book}, it represents a brand new attempt in the context of Transformer-based music modeling. This facilitates extensions of the Transformer, such as the Compressive Transformer \cite{rae20iclr},  Insertion Transformer \cite{pmlr-v97-stern19a}, and Tree Transformer \cite{wang19treetransformer}, to be studied for this task in the future, since the models have now clearer ideas of segment boundaries in music.

Moreover, we further explore  adding other supportive musical tokens capturing higher-level information of music. 
Specifically, to model the expressive rhythmic freedom in music (e.g., tempo rubato), we add a set of \textsc{Tempo} events to allow for local tempo changes per beat.
To have control over the chord progression underlying the music being composed, we introduce the 
\textsc{Chord} events to make the harmonic structure of music explicit, in the same vein of adding the \textsc{Position} and \textsc{Bar} events for rhythmic structure.

\begin{figure}
    \centering
    \includegraphics[width=\linewidth]{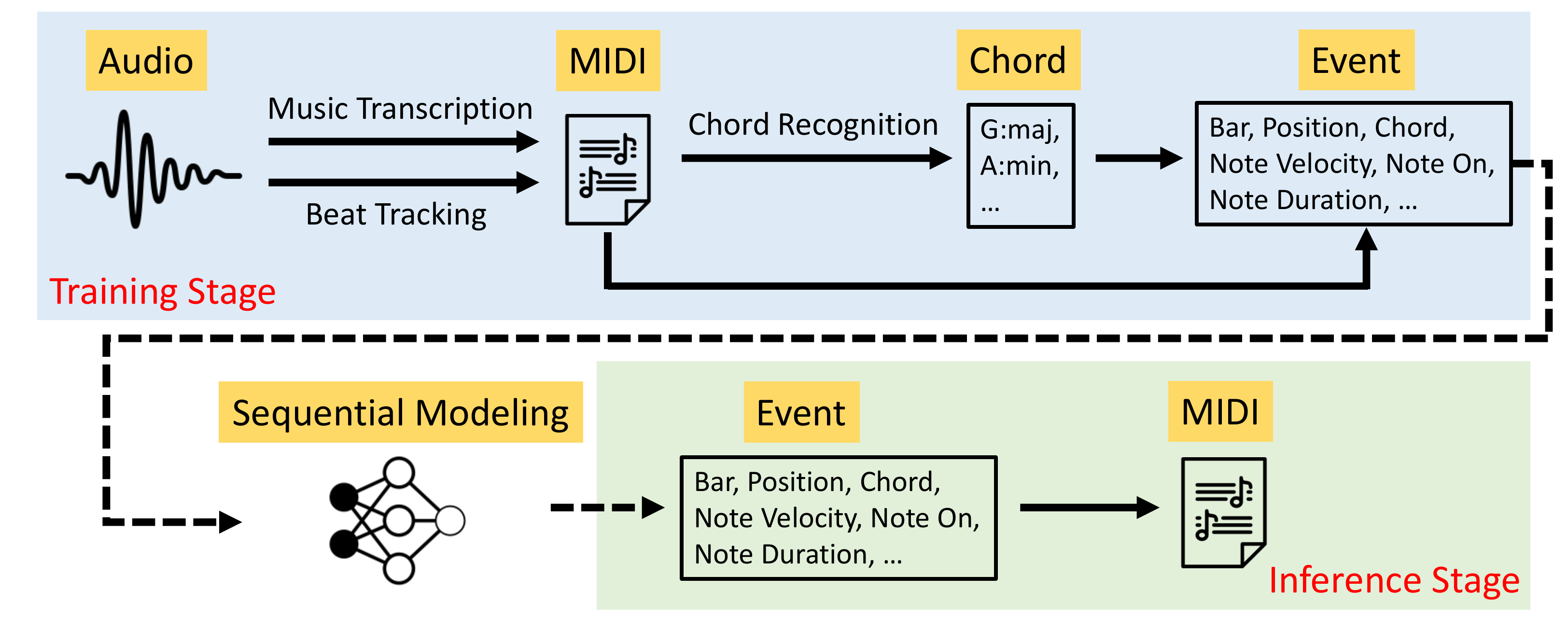}
    \caption{The diagram of the proposed beat-based music modeling and generation framework. The training stage entails the use of music information retrieval (MIR) techniques, such as automatic transcription, to convert an audio signal into an event sequence, which is then fed to a sequence model for training. At inference time, the model generates an original event sequence, which can then be converted into a MIDI file for playback.}
    \label{fig:diagram}
\end{figure}

\begin{figure}[!t]
    \centering
    \includegraphics[width=\linewidth]{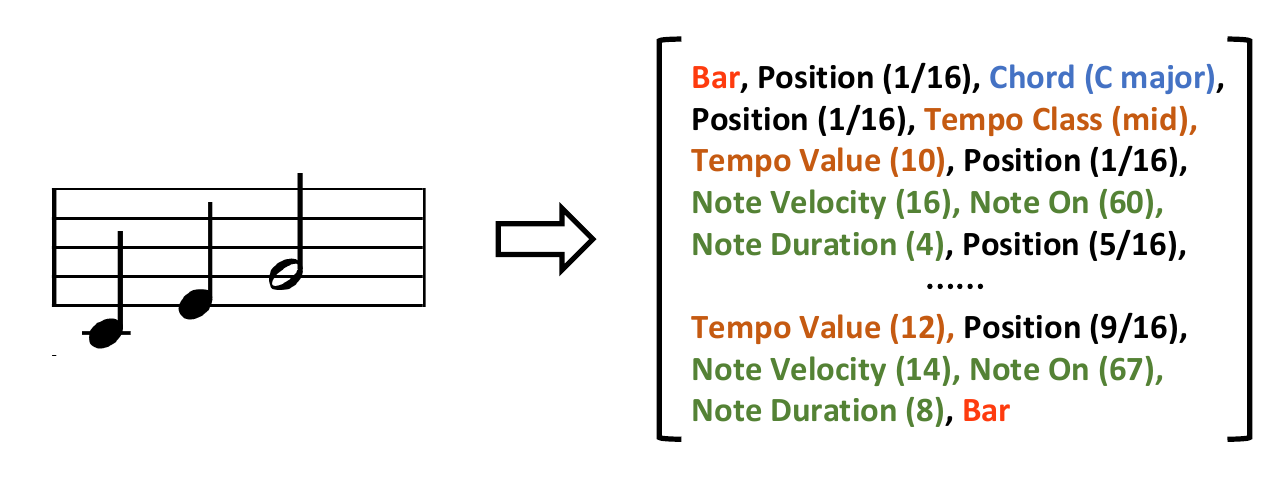}
    \caption{An example of a REMI event sequence and the corresponding music fragment in staff notation.
    The \textsc{Bar}, \textsc{Position} and \textsc{Tempo}-related events entail the use of audio-domain downbeat and beat tracking algorithms \cite{bock2016joint} (see Section \ref{ssc:time-shift-versus-position}), and the \textsc{Chord} events the use of a symbolic-domain chord recognition algorithm (see Section \ref{ssc:chord}).}
    \label{fig:event_representation}
\end{figure}

In achieving all these, the proposed framework for automatic music composition has to integrate audio- and symbolic-domain \textbf{music information retrieval} (MIR) techniques such as downbeat tracking \cite{madmom,bock2016joint,fuentes2018analysis} and chord recognition \cite{fujishima99,scholzEtAl_2008_CochRecoCompChor}, which differentiates it from existing approaches. Moreover, we use automatic music transcription \cite{benetos2018automatic,hawthorne2018onsets,kim2019adversarial} to prepare the MIDI data, so the model learns from \textbf{expressive piano performances}. We show in Figure \ref{fig:diagram} an illustration of the overall framework. Table \ref{tab:event-comparison} compares REMI with the commonly-adopted MIDI-like representation, and Figure \ref{fig:event_representation} gives an example of a REMI event sequence. 

In our experiment (see Section \ref{sc:experiments}), we use Transformer-XL \cite{dai2019transformer} to learn to compose Pop piano music using REMI as the underlying data representation. 
We conduct both objective and subjective evaluation comparing the proposed model against the Music Transformer  \cite{huang2018music}, showing that we are able to improve the state-of-the-art with simple yet major modifications in the data presentation, rather than with sophisticated re-design of the neural network architecture.

For reproducibility, the code, data and pre-trained model are made publicly available.\footnote{\url{https://github.com/YatingMusic/remi}}

\section{Related Work}
\label{sc:related-work}

Recent neural network-based approaches for automatic music composition can be broadly categorized into two groups. \emph{Image-based} approaches such as MidiNet \cite{midinet} and MuseGAN \cite{musegan} use an image-like representation such as the piano roll to represent a score as a matrix of time steps and MIDI pitches, and then use convolution-based operations to generate music. It is easier for such approaches to learn the rhythmic structure of music, as the time steps corresponding to the beats and bar lines are clear in such matrices.
\emph{Language-based} approaches such as the Music Transformer \cite{huang2018music}, on the other hand, may learn the temporal dependency between musical events such as \textsc{Note-On}s better. Yet, due to the limits of the MIDI-like representation outlined in Section \ref{sc:introduction}, existing work may fall short in the rhythmic aspect.
The proposed approach combines the advantage of the image- and language-based approaches by embedding a metrical grid in the event representation.

The idea of designing events for music metadata is similar to CTRL \cite{keskar2019ctrl}, which provided more explicit controls for text generation. However, recent work in neural sequence modeling of music mostly adopt the same MIDI-like event representation proposed by \cite{oore2018time}, or its extensions. For example, extra events denoting the composer, instrumentation and global tempo are used in MuseNet \cite{payne2019musenet} for conditioned generation, and events specifying the \textsc{Note-On} and \textsc{Note-Off} of different instruments are used to achieve multi-instrument music composition  \cite{donahue2019lakhnes}.  However, to our best knowledge, the use of \textsc{Time-Shift} to mark the advance in time has been taken for granted thus far. While this approach has its own merits, we endeavor to explore alternative representations in this paper, again in the specific context of Transformer-based modeling.

\section{New Event-based Representation of Music}
\label{sc:event-based-representation}

In this section, we discuss at length how `REMI' is different from the commonly-adopted `MIDI-like'  representation  (cf. Table \ref{tab:event-comparison}), and how we design the proposed events. 

As discussed in \cite{oore2018time}, a score to be converted into events
can be either a \emph{MIDI score} with no expressive dynamics and timing, or a \emph{MIDI performance} that has been converted from an expressive audio recording by means of, for example, a MIDI keyboard, or automatic music transcription  \cite{benetos2018automatic}. We consider the latter case in this paper, but the discussion below can also be generalized to the former case.

\subsection{\textsc{Note-On} and \textsc{Note Velocity}}
\label{ssc:note-on-and-note-velocity}
The collection of 128 \textsc{Note-On} events indicates the onset of MIDI pitches from 0 (\texttt{C-1}) to 127 (\texttt{G9}), and  \textsc{Note Velocity} indicates the level of dynamics (which correspond to perceptual loudness) of the note event.\footnote{Following \cite{oore2018time}, in our implementation we quantize note velocity into 32 levels, giving rise to 32 different \textsc{Note Velocity} events.} 
Both the MIDI-like and REMI representations have these two types of events.

\subsection{\textsc{Note-Off} versus \textsc{Note Duration}}
\label{ssc:note-off-versus-duration}

In REMI, we use \textsc{Note Duration} events in replacement of the \textsc{Note-Off} events. 
Specifically, we represent each note in a given score with the following three consecutive tokens: a \textsc{Note Velocity} event, a \textsc{Note-On} event, and a \textsc{Note Duration} event.
There are advantages in doing so:
\begin{itemize}
    \item In MIDI-like, the duration of a note has to be  inferred from the time gap between a \textsc{Note-On}  and the corresponding \textsc{Note-Off}, by accumulating the time span of the \textsc{Time Shift} events in between. In REMI, note duration is made explicit, facilitating modeling the rhythm of music.
    \item In  MIDI-like, a \textsc{Note-On} event and the corresponding \textsc{Note-Off} are usually several events apart.\footnote{For example, in our implementation, there are on average 21.7$\pm$15.3 events between a pair of \textsc{Note-On} and \textsc{Note-Off}, when we adopt the MIDI-like event representation.} As a result, a sequence model may find it difficult to learn that \textsc{Note-On} and \textsc{Note-Off} must appear in pairs, generating dangling \textsc{Note-On} events without the corresponding \textsc{Note-Off}. 
    We would then have to use some heuristics (e.g., maximal note duration) to turn off a note in post-processing. 
    With REMI, we are free of such an issue, because the sequence model can easily learn from the training data that a \textsc{Note Velocity} event has to be followed by a \textsc{Note-On} event, and then right away a \textsc{Note Duration} event.
\end{itemize}

\subsection{\textsc{Time-Shift} versus \textsc{Position} \& \textsc{Bar}}
\label{ssc:time-shift-versus-position}

Empirically, we find that it is not easy for a  model to generate music with steady beats using the \textsc{Time-Shift} events. When listening to the music generated, the intended bar lines drift over time and the rhythm feels unstable. We attribute this to the absence of a metrical structure in the MIDI-like representation.
Mistakes in \textsc{Time-Shift} would lead to accumulative error of timing in the inference phase, which is not obvious in the training phase due to the common use of teacher forcing strategy \cite{doya1992bifurcations} in training recurrent models.

To address this issue, we propose to use the combination of \textsc{Bar} and \textsc{Position} events instead. Both of them are readily available from the musical scores, or from the result of automatic music transcription (with the help of a beat and downbeat tracker \cite{bock2016joint}); they are simply discarded in the MIDI-like representation. 
While \textsc{Bar} marks the bar lines, \textsc{Position} points to one of the $Q$ possible discrete locations in a bar. 
Here $Q$ is an integer that denotes the time resolution adopted to represent a bar. For example, if we consider a 16-th note time grid as \cite{roberts2018hierarchical,huang2018music}, $Q=16$.
Adding \textsc{Bar} and \textsc{Position} therefore provides a metrical context for models to ``count the beats'' and to compose music bar-after-bar.\footnote{We find in our implementation that models learn the meaning of \textsc{Bar} and \textsc{Position}  quickly---right after a few epochs the model knows that, e.g., \textsc{Position (9/$Q$)} cannot go before \textsc{Position (3/$Q$)}, unless there is a \textsc{Bar} in between. No postprocessing is needed to correct errors. See Figure \ref{fig:loss} for the learning curve seen in our implementation.}\footnote{We note that the idea of using \textsc{Bar} and \textsc{Position} has been independently proposed before \cite{genchel19arxiv}, though in the context of RNN-based not Transformer-based models.
}

We note that there are extra benefits in using  \textsc{Position} \& \textsc{Bar}, including 1) we can  more easily learn the dependency (e.g., repetition) of note events occurring at the same \textsc{Position ($\star$/$Q$)} across bars; 2) we can add bar-level conditions to condition the generation process if we want; 3) we have time reference to coordinate the generation of different tracks for the case of multi-instrument music.

\subsection{\textsc{Tempo}}
\label{ssc:tempo-changes}
In an expressive musical performance, the temporal length (in seconds) of each bar may not be the same. To account for such local changes in tempo (i.e., beats per minute; BPM), we add \textsc{Tempo} events every beat (i.e., at \textsc{Position (1/$Q$)}, \textsc{Position ($\big(\frac{Q}{4}+1\big)$/$Q$)}, etc). In this way, we have a flexible time grid for expressive rhythm.\footnote{As exemplified in Figure \ref{fig:event_representation}, we use a combination of  \textsc{Tempo Class} events (low, mid, high) and \textsc{Tempo Value} events to represent local tempo values of 30--209 BPM.}  Unlike the \textsc{Position} events, information regrading the \textsc{Tempo} events may not be always available in a MIDI score. But, for the case of MIDI performances, one can derive such \textsc{Tempo} events with the use of an audio-domain tempo estimation function \cite{bock2016joint}.

\subsection{\textsc{Chord}}
\label{ssc:chord}

As another set of supportive musical tokens,
we propose to encode the chord information into input events. Specifically, chords are defined as any harmonic set of pitches consisting of multiple notes sounding together or one after another. A chord consists of a  root note and a chord quality \cite{mcfee2017structured}.
Here, we consider 12 chord roots (\texttt{C,C\#,D,D\#,E,F,F\#,G,G\#,A,A\#,B}) and five chord qualities (major, minor, diminished, augmented, dominant), yielding 60 possible \textsc{Chord} events, covering the triad and seventh chords. However, the set of \textsc{Chord} events can be further expanded as long as there is a way to get these chords from a MIDI score or a MIDI performance in the data preparation process. 
We note that the \textsc{Chord} events are just ``symbols''--- the notes are still generated with the \textsc{Note-on} events after them.

Following the time grid of REMI, each \textsc{Tempo} or \textsc{Chord} event is preceded by a \textsc{Position} event.

We note that music composed by a model using the MIDI-like representation also exhibits the use of chords, as such note combinations can be found in the training data by the sequence model itself. However, by explicitly generating \textsc{Tempo} and \textsc{Chord} events, tempo and chord become controllable.


\section{Proposed Framework}

Again, a diagram of the proposed beat-based music modeling and generation framework can be found in Figure \ref{fig:diagram}. We provide details of some of the major components below.

\subsection{Backbone Sequence Model}
\label{sc:backbone-model}

The Transformer \cite{vaswani2017attention} is a neural sequence model that uses  self-attention  to bias its prediction of the current token based on a subset of the past tokens.
This design has been shown effective for modeling the structure of music.
For example, with the help of a relative-positional encoding method \cite{shaw2018self}, Music Transformer  is claimed in \cite{huang2018music} to be able to compose expressive classical piano music with a coherent structure of up to one minute. 

Transformer-XL \cite{dai2019transformer} extends Transformer by introducing the notion of recurrence and revising the positional encoding scheme.
The recurrence mechanism enables the model to leverage the information of past tokens beyond the current training segment, thereby looking further into the history.
Theoretically, Transformer-XL can encode arbitrarily long context into a fixed-length representation. Therefore, we adopt Transformer-XL as our backbone model architecture.\footnote{We have implemented both Transformer- and Transformer-XL based models and found the latter composes Pop piano music with better temporal coherence perceptually. However, as the use of Transformer-XL for automatic music composition has been reported elsewhere \cite{donahue2019lakhnes}, we omit such an empirical performance comparison between the XL and non-XL versions.}

In the training process, each input sequence is divided into ``segments'' of a specific length (set to 512 events in our implementation). Given such segments, the overall computational procedure for an $N$-layer Transformer-XL with $M$ heads can be summarized as:
\begin{align}
    \tilde{\mathbf{h}}^{n-1}_{\tau} &= [\text{stop\_gradient}(\mathbf{h}^{n-1}_{\tau -1})\circ \mathbf{h}^{n-1}_{\tau}] \,, \\
    \mathbf{q}^{n}_{\tau},~\mathbf{k}^{n}_{\tau},~\mathbf{v}^{n}_{\tau} &= \mathbf{h}^{n-1}_{\tau}\mathbf{W}^{\top}_{q},~\tilde{\mathbf{h}}^{n-1}_{\tau}\mathbf{W}^{\top}_{k},~\tilde{\mathbf{h}}^{n-1}_{\tau}\mathbf{W}^{\top}_{v} \,, \\
    \mathbf{a}^{n}_{\tau ,i} &= \text{masked\_softmax}(\mathbf{q}^{n\top}_{\tau ,i}\mathbf{k}^{n}_{\tau ,i}+\mathbf{R})\mathbf{v}^{n}_{\tau ,i} \,, \\
    \mathbf{a}^{n}_{\tau} &= [\mathbf{a}^{n}_{\tau ,1}\circ \mathbf{a}^{n}_{\tau ,2} \circ...\circ \mathbf{a}^{n}_{\tau ,m}]^{\top}\mathbf{W}^{n}_{o} \,, \\
    \mathbf{o}^{n}_{\tau} &= \text{layernorm}(\mathbf{a}^{n}_{\tau}+\mathbf{h}^{n-1}_{\tau}) \,, \\
    \mathbf{h}^{n}_{\tau} &= \text{max}(0, \mathbf{o}^{n}_{\tau}\mathbf{W}^{n}_{1}+\mathbf{b}^{n}_{1})\mathbf{W}^{n}_{2}+\mathbf{b}^{n}_{2}  \,,
\end{align}
where $\mathbf{h}^{n}_{\tau}$ denotes the $n$-th layer hidden features for the $\tau$-th segment, $\mathbf{R}$ denotes the relative positional encodings designed for the Transformer-XL \cite{dai2019transformer}, $\mathbf{a}^{n}_{\tau,i}$ indicates the attention features from the $i$-th head, and $\mathbf{q}$, $\mathbf{k}$, $\mathbf{v}$ denote the query, key and values in the computation of self-attention \cite{vaswani2017attention}, respectively.
The main difference between Transformer and Transformer-XL lies in the usage of the features from the segments prior to the current segment (i.e., the segments are from the same input MIDI sequence of a music piece) when updating the model parameters based on that current segment. This mechanism brings in longer temporal context and speeds up both the training and inference processes.


Transformer-like sequence models learn the dependency among elements (i.e., events here) of a sequence. Therefore, although we have quite a diverse set of events featuring different characteristics in the proposed REMI representation, we opt for the simple approach of using a single Transformer-XL to model all these events at once. In other words, we do not consider the alternative, possibly more computationally intensive, approach of establishing multiple Transformer models, one for a non-overlapping subset of the events, followed by some mechanisms to communicate between them.

\begin{table}[!t]
\centering
\begin{tabular}{@{}l|llll@{}}
\toprule
Chord & Required & \begin{tabular}[c]{@{}l@{}}Gain\\ 1 point\end{tabular} & \begin{tabular}[c]{@{}l@{}}Deduct\\ 1 point\end{tabular} & \begin{tabular}[c]{@{}l@{}}Deduct\\ 2 points\end{tabular} \\ \midrule
Major & 0, 4 & 7 & 2, 5, 9 & 1, 3, 6, 8, 10 \\
Minor & 0, 3 & 7 & 2, 5, 8 & 1, 4, 6, 9, 11 \\
Diminished & 0, 3, 6 & 9 & 2, 5, 10 & 1, 4, 7, 8, 11 \\
Augmented & 0, 4, 8 & - & 2, 5, 9 & 1, 3, 6, 7, 10 \\
Dominant & 0, 4 ,7, 10 & - & 2, 5, 9 & 1, 3, 6, 8, 11 \\ \bottomrule
\end{tabular}
\caption{The proposed rule-based scoring criteria for establishing the \textsc{Chord} events via pitch intervals in the chromatic scale.}
\label{tab:chord-recognition}
\end{table}

\subsection{Beat Tracking and Downbeat Tracking}
\label{sc:beat_tracking}

To create the \textsc{Bar} events, we employ the recurrent neural network model proposed by \cite{bock2016joint} to estimate from the audio files the position of the `downbeats,' which correspond to the first beat in each bar.
The same model is used to track the beat positions to create the  \textsc{Tempo} events \cite{madmom}.
We obtain the tick positions between beats by linear interpolation, and then align the note onsets and offsets to the nearest tick.
To eliminate the imprecision of transcription result, we further quantize the onset and offset times according to the assumed time grid (e.g., to the 16-th note when $Q=16$).

\subsection{Chord Recognition}
\label{sc:chord_model}

We establish the 60 \textsc{Chord} events described in Section \ref{ssc:chord} by designing a heuristic rule-based symbolic-domain chord recognition algorithm to the transcribed MIDI files.
First, we compute binary-valued ``chroma features'' \cite{fujishima99} for each tick, to represent the activity of 12 different pitch classes ignoring the pitch's octave. 
Then, we use a sliding window to assign ``likelihood scores'' to every active note for each 2-beat and 4-beat segment.
After summarizing the chroma features of the current segment, 
we consider every note in a segment as a candidate root note of the \textsc{Chord} of that segment, and calculate its pitch intervals to all the other notes in that segment.
The look-up table shown in Table \ref{tab:chord-recognition} is employed to assign likelihood scores to a pair of root note and chord quality based on the pitch intervals. 
Each chord quality has its required set of pitch intervals, and scoring functions.
Finally, we recursively label the segments by the \textsc{Chord} symbol with the highest likelihood score.

\begin{figure}
\centering
\begin{subfigure}{0.45\linewidth}
    \centering
    \includegraphics[width=\linewidth]{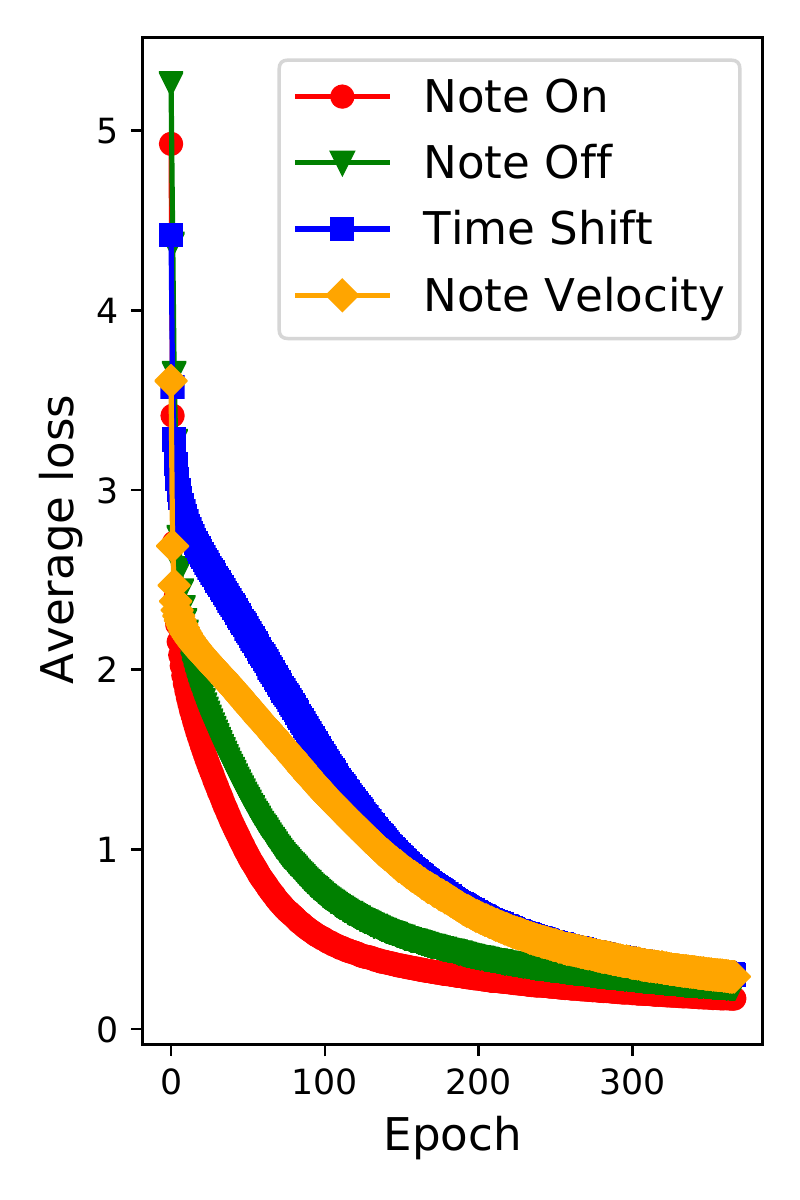}
    \caption{Baseline 1}
    \label{fig:loss-on-off-ts}
\end{subfigure}
\begin{subfigure}{0.45\linewidth}
    \centering
    \includegraphics[width=\linewidth]{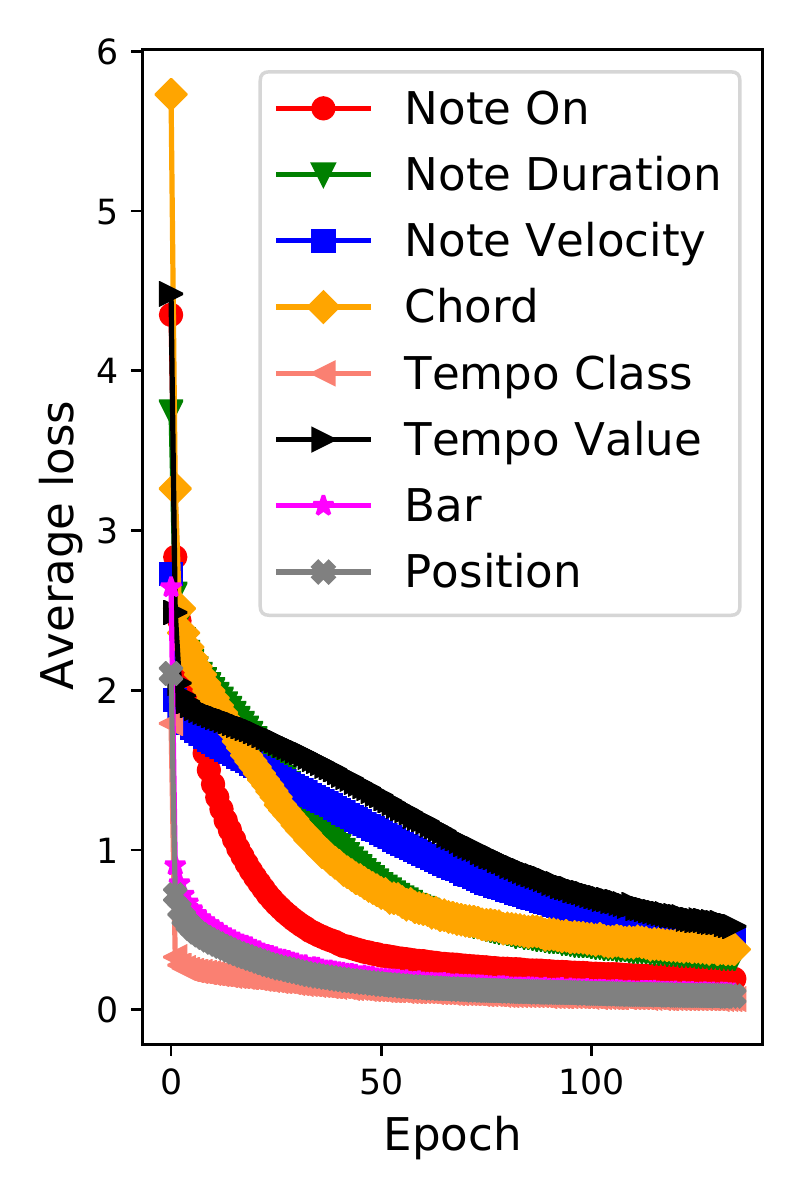}
    \caption{REMI}
    \label{fig:loss-on-duration-position-tempo-chord}
\end{subfigure}
\caption{Average cross-entropy loss for different types of events as the training process proceeds. (a) An adapted version of the Music Transformer \cite{huang2018music}; (b) the proposed model.}
\label{fig:loss}
\end{figure}

\begin{table*}[!t]
\centering
\begin{tabular}{@{}l|llcc|ccc@{}}
\toprule
\multirow{2}{*}{Method} & \multirow{2}{*}{Note offset} & \multirow{2}{*}{Time grid} & \multirow{2}{*}{\textsc{Tempo}}  & \multirow{2}{*}{\textsc{Chord}} & Beat & Downbeat& Downbeat \\
&  &  &  & & STD& STD& salience \\ \midrule
Baseline 1 & \textsc{Note-Off} & \textsc{Time-Shift} (10-1000ms) & & & 0.0968 & 0.3561 & 0.1033 \\
Baseline 2 & \textsc{Duration} & \textsc{Time-Shift} (10-1000ms) & & & 0.0394 & 0.1372 & 0.1651 \\
Baseline 3 & \textsc{Duration} & \textsc{Time-Shift} (16th-note multiples) & & & \textbf{0.0396} & \textbf{0.1383} & 0.1702 \\
\midrule
\multirow{4}{*}{REMI} & \textsc{Duration} & \textsc{Position} \& \textsc{Bar} & \cmark & \cmark & \textbf{0.0386} & \textbf{0.1376} & 0.2279 \\
 & \textsc{Duration} & \textsc{Position} \& \textsc{Bar} & \cmark & & 0.0363 & 0.1265 & \textbf{0.1936} \\
 & \textsc{Duration} & \textsc{Position} \& \textsc{Bar} & & \cmark & 0.0292 & 0.0932 & 0.1742 \\ 
 & \textsc{Duration} & \textsc{Position} \& \textsc{Bar} & & & 0.0199 & 0.0595 & \textbf{0.1880} \\ \midrule
Training data~~~ & & & & & 0.0607 & 0.2163 & 0.2055 \\ \bottomrule
\end{tabular}
\caption{Quantitative comparison of different models for Pop piano composition, evaluating how the composed music exhibit rhythmic structures. 
We report the average result across songs.
For all the three objective metrics, the values are the closer to those of the `Training data' shown in the last row the better---the training data are the transcribed and 16-th note quantized version of the 775 Pop songs.  `Baseline 1' represents a Transformer-XL version of the Music Transformer, adopting a MIDI-like event representation. The other two baselines are its improved variants. We highlight the two values closest to those of the training data in bold.}
\label{tab:objective-evaluation}
\end{table*}

\begin{figure*}[!t]
    \centering
    \includegraphics[width=\linewidth]{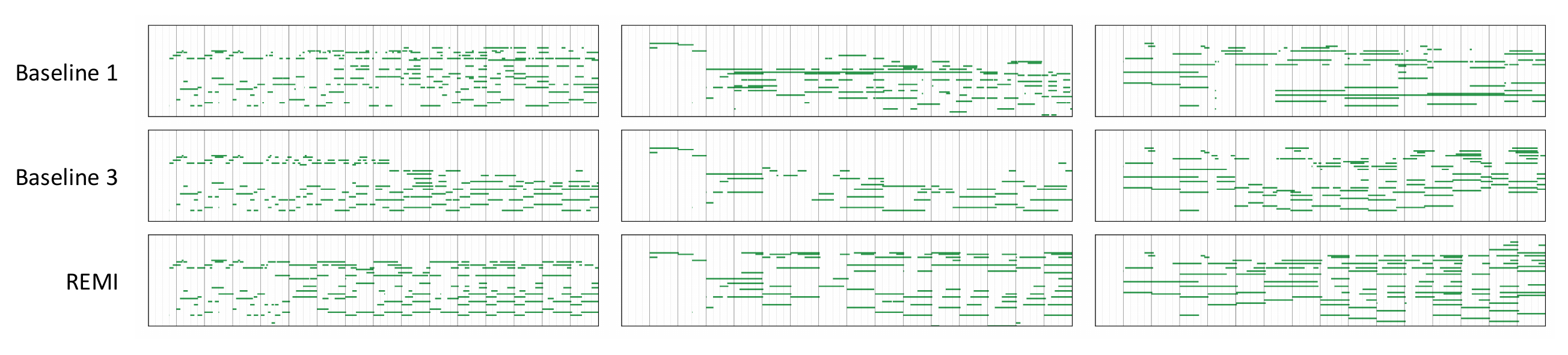}
    \caption{Examples of the generated piano rolls of different models.  
    These are 12 bars generated to continue a given prompt of 4-bar human piano performance.  We show the result for three different prompts (from left to right). The thicker vertical lines indicate the bar lines; best viewed in color.}
    \label{fig:generated_samples}
\end{figure*}

\section{Evaluation}
\label{sc:experiments}

We report both objective and subjective evaluations aiming to validate the effectiveness of the proposed model over the Music Transformer \cite{huang2018music} for the case of composing expressive Pop piano music. In what follows, we present the training data we used to train both our model and variants of the the Music Transformer, then some implementation details, and finally the performance study.

\subsection{Dataset}
\label{sc:db}

We intend to evaluate the effectiveness of the proposed approach for Pop piano composition, as Pop is a musical genre that features salient rhythmic structures.  
In doing so, we collect audio files of Pop piano music from the Internet.
A total number of 775 pieces of piano music played by different people is collected, amounting to approximately 48 hours' worth of data. They are covers\footnote{A `cover' is a new recording by someone other than the original artist or composer of a commercially released song.} of various Japanese anime, Korean popular and Western popular songs, playing only with the piano.
We then apply ``Onsets and Frames'' \cite{hawthorne2018onsets}, the state-of-the-art approach for automatic piano transcription, to estimate the pitch, onset time, offset time and velocity of the musical notes of each song, converting the audio recordings into MIDI performances.

We select the Pop piano covers such that they are all in 4/4 time signature, simplifying the modeling task. Accordingly, a bar is composed of four beats.
Furthermore, following \cite{huang2018music}, we consider the 16-th note time grid and quantize each bar into $Q=16$ intervals. 

In our preliminary experiments, we have attempted to use a finer time grid (e.g., 32-th or 64-th note) to reduce quantization errors and to improve the expression of music (e.g., to include triplets, swing, mirco-timing variations, etc). However, we found that this seems to go beyond the capability of the adopted Transformer-XL architecture. The generated compositions using a finer time grid tends to be fragmented and not pleasant to listen to. This is an issue that has to be addressed in the future. We stick with $Q=16$ hereafter.

Due to copyright restrictions, we plan to make the training data publicly available not as audio files but as the transcribed MIDI files and the converted REMI event sequences.

\subsection{Baselines \& Model Settings}
\label{ssc:baselines}
We consider three variants of the Music Transformer \cite{huang2018music} as the baselines in our evaluation. 
The first one, dubbed \textbf{Baseline 1}, follows fairly faithfully the model settings of \cite{huang2018music}, except that we use Transformer-XL instead of Transformer (for fair comparison with our model).
The other two differ from the first one only in the adopted event representation.
While Baseline 1 employs exactly the MIDI-like representation,
\textbf{Baseline 2} replaces \textsc{Note-Off} by \textsc{Note Duration}, and \textbf{Baseline 3} further 
modifies the time steps taken in \textsc{Time-Shift} from multiples of 10ms to multiples of the 16-th note.
In this way, Baseline 3 has a time grid similar to that of REMI, making Baseline 3 a strong baseline.

%
For either the baselines or the proposed model adopting the REMI representation, we train a single Transformer-XL with $N=12$ self-attention layers and $M=8$ attention heads.
The length of the training input events (i.e., the segment length) and the recurrence length (i.e., the length of the segment ``cached'' \cite{dai2019transformer}) are both set to 512.
The total number of learnable parameters is approximately 41M.
Training a model with an NVIDIA V100 with mini-batch size of 16 till the training loss (i.e., cross-entropy) reaches a certain level takes about 9 hours. 
We find that Baseline 1 needs a smaller learning rate and hence longer training time.
For processing the MIDI files, we use the miditoolkit. \footnote{\url{https://github.com/YatingMusic/miditoolkit}}

Figure \ref{fig:loss} shows the training loss of different event types as the training unfolds. Figure \ref{fig:loss-on-off-ts} shows that Baseline 1 struggles the most for learning \textsc{Time-Shift}. Moreover, \textsc{Note-Off} has higher loss than \textsc{Note-On}, suggesting that they are not recognized as event pairs. 
In contrast, Figure \ref{fig:loss-on-duration-position-tempo-chord} shows that the \textsc{Position} events in REMI are easy to learn, facilitating learning the rhythm of music.\footnote{According to \cite{hawthorne2018onsets,kim2019adversarial}, piano transcription models do better in estimating note onsets and pitches, than offsets and velocities. This may be why our model has higher loss for \textsc{Note Duration} and \textsc{Note Velocity}.}

\subsection{Objective Evaluation}
\label{ssc:objective-evaluation}

The objective evaluation assesses the rhythmic structure of the generated compositions. 
Specifically, we employ each model to randomly generate 1,000 sequences, each with 4,096 events,
using the temperature-controlled stochastic sampling method with top-\emph{k} \cite{keskar2019ctrl}. 
We convert the generated sequences into MIDI and then render them into audio via a piano synthesizer.\footnote{\url{https://github.com/YatingMusic/ReaRender}} Then, we apply the joint beat and downbeat tracking model of \cite{bock2016joint} to the resulting audio clips.\footnote{We choose to use the model of \cite{bock2016joint} for it achieved state-of-the-art performance for beat and downbeat tracking for a variety of musical genres, especially for Pop music.} 

The model \cite{bock2016joint} has two components.
The first one estimates the probability, or \emph{salience}, of observing beats and downbeats for each time frame via a recurrent neural network (RNN). The second one applies a dynamic Bayesian network (DBN) to the output of the RNN to make binary decisions of the occurrence of beats and downbeats. We use the output of the RNN to calculate the downbeat salience, and the output of the DBN for the beat STD and downbeat STD.
Specifically, given a piece of audio signal, the model \cite{bock2016joint} returns a time-ordered series of ($\tau_t^\text{B}$, $s_t^\text{B}$), where $\tau_t^\text{B}$ denotes the estimate time (in seconds) for the $t$-th beat by the DBN, and $s_t^\text{B}\in [0,1]$ the salience associated with that beat estimated by the RNN. It also returns similarly a series of ($\tau_t^\text{D}$, $s_t^\text{D}$) for the downbeats; see Figure \ref{fig:salience} for examples of such series. 
From the tracking results, we calculate the following values for each audio clip:
\begin{itemize}
    \item \textbf{Beat STD}:  
    the standard deviation of $(\tau_t^\text{B}-\tau_{t-1}^\text{B})$, which is always positive, over the beat estimates for that clip.
    \item \textbf{Downbeat STD}: similarly the standard deviation of $(\tau_t^\text{D}-\tau_{t-1}^\text{D})$  over the downbeat estimates for that clip. Both beat STD and downbeat STD assess the consistency of the rhythm. 
    \item \textbf{Downbeat salience}: the mean of $s_t^\text{D}$ over the downbeat estimates for that clip, indicating the salience of the rhythm.
\end{itemize}
We report the average values across the audio clips, assuming that these values are closer to the values calculated from the training data (which have been quantized to the 16-th note grid) the better.\footnote{Small beat/downbeat STD implies that the music is too rigid, whereas large STD implies that the rhythm is too random. We want the rhythm to be stable yet flexible. }

Tables \ref{tab:objective-evaluation} shows the result of the baselines and a few variants (ablated versions) of the REMI-based model.
From Beat STD and Downbeat STD, we see that the result of Baseline 1, which resembles Music Transformer \cite{huang2018music}, features fairly inconsistent rhythm, echoing the example shown in Figure \ref{fig:salience:MIDI}.
The STD is lower when \textsc{Note Duration} is used in place of \textsc{Note-Off}, highlighting the effect of \textsc{Note Duration} in stabilizing the rhythm.
Interestingly, we see that the REMI models without the \textsc{Tempo} events have much lower STD than the training data, suggesting that \textsc{Tempo} is important for expressive rhythmic freedom.

From downbeat salience, the REMI models outnumber all the baselines, suggesting the effectiveness of \textsc{Position} \& \textsc{Bar}.
Moreover, the gap between the result of Baseline 1 and Baseline 2 further supports the use of the \textsc{Note Duration} events.

\begin{table}[!t]
\centering
\begin{tabular}{@{}l|llll@{}}
\toprule
& Baseline 1 & Baseline 3 & REMI \\ \midrule
`pros' group & 1.77 & 2.03 & 2.20 \\
`non-pros' group & 1.66 & 1.90 & 2.44 \\ \midrule
all participants & 1.73 & 1.98 & 2.28 \\
\bottomrule
\end{tabular}
\caption{The average scores for the subjective preference test on a three-point scale from 1 (like the least) to 3 (like the most).}
\label{tab:subject-evaluation}
\end{table}
\begin{table}[!t]
\centering
\begin{tabular}{@{}ll|lll@{}}
\toprule
Pairs & & Wins & Losses & \emph{p}-value \\ \midrule
REMI & Baseline 1 & 103 & 49 & 5.623e-5 \\
REMI & Baseline 3 & 92 & 60 & 0.0187 \\
Baseline 3 & Baseline 1 & 90 & 62 & 0.0440 \\ \bottomrule
\end{tabular}
\caption{The result of pairwise comparison from the user study, with the \emph{p}-value of the Wilcoxon signed-rank test.}
\label{tab:wilcoxon}
\end{table}

\subsection{Subjective Evaluation}
\label{ssc:subjective-evaluation}
The best way to evaluate a music composition model remains today to be via a listening test. 
To have a common ground to compare different models, we ask the models to \emph{continue} the same given prompt.
For doing so, we prepare an additional set of 100 Pop piano performances (that have no overlaps with the training set), process them according to the procedures described in Section \ref{sc:db}, and take the first four bars from each of them as the prompts.
The following three models are asked to generate 16 bars continuing each prompt, and got evaluated in an online listening test: `Baseline 1,' `Baseline 3' and `REMI without \textsc{Chord}.' We do not use \textsc{Chord} here to make the rhythmic aspect the major point of difference among the models.

We distribute the listening test over our social circles globally and solicit the response from 76 participants.
51 of them understand basic music theory and have the experience of being an amateur musician, so we consider them as professionals (denoted as `pros'). A participant has to listen to two randomly picked sets of samples and evaluate, for each set, which sample they \emph{like the most} and \emph{the least}, in any evaluation criteria of their choice.
Each set contains the result of the three models (in random order) for a given prompt.

Table \ref{tab:subject-evaluation} shows the aggregated scores, and Table \ref{tab:wilcoxon} the result of broken-down pairwise comparisons, along with the $p$-value of the Wilcoxon signed-rank test.
We see that REMI is preferred by both pros and non-pros, and that the difference between REMI and Baseline 1 is significant ($p<0.01$).

Figure \ref{fig:generated_samples} provides examples of the generated continuations. 
From the optional verbal feedbacks of the participants,  the music generated by REMI are perceptually more pleasing and are more in line with the prompt.
Audio examples 
can be found at our demo website.\footnote{\url{https://drive.google.com/drive/folders/1LzPBjHPip4S0CBOLquk5CNapvXSfys54}}

\subsection{Controllable \textsc{Chord} and \textsc{Tempo}}
\label{ssc:controllable}
\begin{figure}[!t]
\centering
\begin{subfigure}[t]{.95\linewidth}
    \centering
    \includegraphics[width=\linewidth]{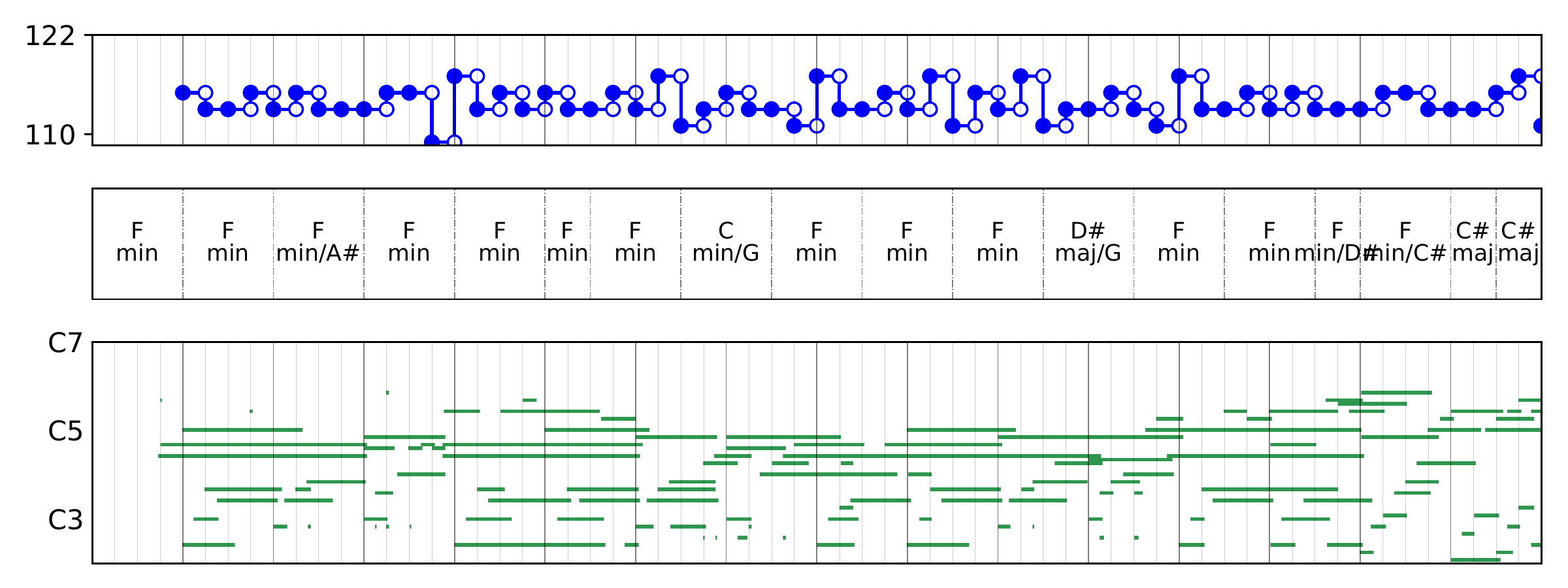}
    \caption{A human-made musical piece containing the prompt used below}
    \label{fig:c-original}
\end{subfigure}
\par\medskip
\begin{subfigure}[t]{.95\linewidth}
    \centering
    \includegraphics[width=\linewidth]{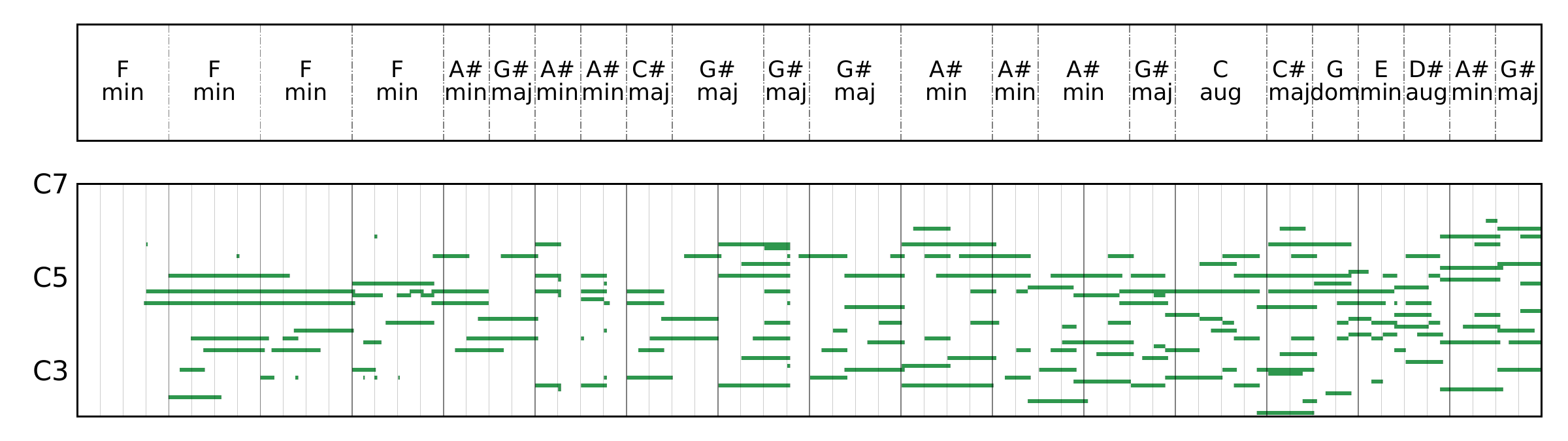}
    \caption{The result when we enforce no  F:minor chord after the 4th bar}
    \label{fig:c-chord}
\end{subfigure}
\par\medskip
\begin{subfigure}[t]{.95\linewidth}
    \centering
    \includegraphics[width=\linewidth]{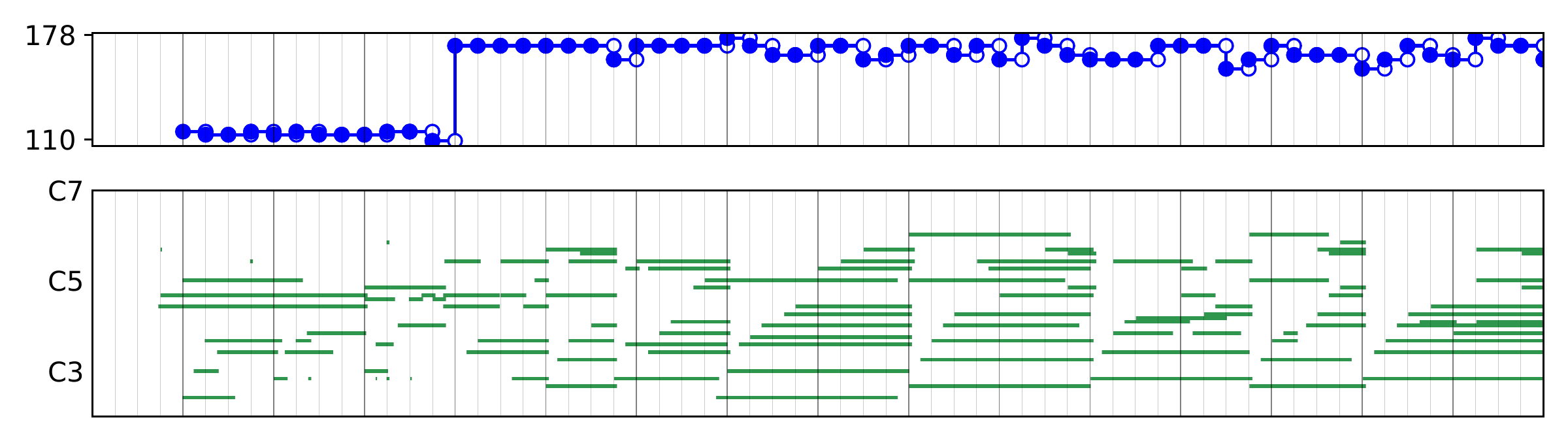}
    \caption{The result when we enforce using \textsc{Tempo Class (high)}}
    \label{fig:c-tempo}
\end{subfigure}
\caption{An example illustrating how we can condition the \textsc{Tempo} and \textsc{Chord} events (the first four bars is the prompt).}
\label{fig:controllable}
\end{figure}

Finally, we demonstrate the controllability of  \textsc{Tempo} and \textsc{Chord} of our model.
To achieve this, we can simply force the model not to generate specific events by masking out the corresponding probabilities of the model output.
Figure \ref{fig:controllable} shows the piano rolls, (optionally) the \textsc{Tempo} and  \textsc{Chord} events generated by our model under different conditions.
Figure \ref{fig:c-chord} shows that the model selects chords that are harmonically close to F:minor when we prohibit it from generating F:minor. 
Figure \ref{fig:c-tempo} shows controlling the \textsc{Tempo Class}  affects not only the  \textsc{Tempo Value} but also the note events.

\section{Conclusion}
\label{sc:conclusion}
In this paper, we have presented REMI, a novel MIDI-derived event representation for sequence model-based composition of expressive piano music.
Using REMI as the underlying data representation, we have also built a Pop Music Transformer that outperforms that state-of-the-art Music Transformer for composing expressive Pop piano music, as validated in objective and subjective studies. 
We take this as a proof-of-concept that it is beneficial to embed prior human knowledge of music through including components of music information retrieval (MIR) techniques, such as downbeat estimation and chord recognition, to the overall generative framework.
For future work, we plan to incorporate other high-level information of music, such as grooving \cite{fruhauf2013music} and musical emotion \cite{yang2011music}, 
to learn to compose additional tracks such as the guitar, bass, and drums, to further improve the resolution of the time grid, to condition the generative process with other multimedia input such as the lyrics \cite{yu2019conditional} or a video clip \cite{yu12mm}, and also to study other model architectures that can model longer sequences (e.g., \cite{rae20iclr}) and thereby learn better the long-term structure in music.

\section{Acknowledgement}\label{sec:ack}
The authors would like to thank \textbf{Wen-Yi Hsiao} (Taiwan AI Labs) for helping with processing the MIDI data and with rendering the MIDIs to audios for subjective study.

\bibliographystyle{ACM-Reference-Format}
\bibliography{remy}

\end{document}